\documentclass[reprint,aps,pre,amsmath,amssymb,superscriptaddress,showpacs]{revtex4-1}
\usepackage{graphicx}
\usepackage{hyperref}
\usepackage{color}

\begin{document}

\title{Local network evolution rules driving shortest path multiplicity}

\author{Alexei Vazquez}
\email{alexei@nodeslinks.com}
\affiliation{Nodes \& Links Ltd, Salisbury House, Station Road, Cambridge, CB1 2LA, UK}

\begin{abstract}
The shortest path multiplicity, here denoted by $\mu$, is an important metric of complex networks. For real networks $\mu$ is high and it correlates with the network community structure. Since local network evolution induces network communities, it is possible that a high shortest path multiplicity is the natural expectation of local evolution rules. Here I demonstrate, by means of numerical simulations and exact results, that this is indeed the case. For random graphs with arbitrary degree distributions $p_k$, $\langle\mu\rangle\sim \langle k(k-1)\rangle / (\langle k\rangle e)$, growing with the network size when $p_k\sim k^{-\gamma}$ and $\gamma\leq3$. For networks generated by stochastic local rules, $\langle\mu\rangle$ increases with the network size and it does so faster than what is observed in their randomized versions. Furthermore, the number of communities increases with the network size and the correlation with $\langle \mu\rangle$ follows.
\end{abstract}

\maketitle

\section{Introduction}

The fundamental task of network science is to determine the minimum set of properties and models to explain what we observe in real networks. New discoveries should be subject to the scrutiny of existing concepts. That is the case of the recent observation by Deng {\em et al} \cite{deng2026multipath}, a correlation between the number of communities and the density of multiple shortest paths in real networks. This correlation could be the consequence of an upstream factor that causes both features. And it turns out there is an obvious candidate: local evolution rules.

Local network evolution rules are node/link addition mechanisms of network growth \cite{vazquez03local}. Local rules are inspired by the natural evolution of real networks. Web pages are created by copying other pages \cite{kleinberg99}. We cite references that we found reading another publication \cite{vazquez01rs, vazquez01citation}. Friends of friends become friends \cite{growing_social_networks_jin01, communities_bianconi_2014}.  Proteins increase their connectivity when their interacting partners are duplicated \cite{vazquez03dup, pastor-satorras03, chung03, copying_krapivsky_2005}. Local evolution rules lead to the emergence of network communities \cite{growing_social_networks_jin01, sole_spontaneous_2007, communities_bianconi_2014}: beyond a certain network size we can warrant that the network will exhibit communities \cite{vazquez2025local}. These communities are not written in the network evolution rules. Yet they are imputed by the state of the art methods to detect network communities.

Local evolution rules act on the shortest path multiplicity through two channels at once, and both are needed. First, they are inherently preferential. A node is the more likely to be duplicated, visited by a random walk, or found at the end of a randomly chosen link, the more links it already has. Preferential attachment generates broad degree distributions, and a broad degree distribution by itself already makes the average shortest path multiplicity grow with the network size, as shown in Sec.~\ref{random}. Second, local rules close short cycles, and cycles create alternative geodesics that the degree sequence alone does not account for. I therefore raise the hypothesis
\[
\mbox{local rules} \Rightarrow \mbox{preferential attachment} + \mbox{cycles} \Rightarrow \mu,
\]
that is, the high shortest path multiplicity of real networks is rooted on their local growth dynamics, through both channels. Furthermore, since those local rules induce the formation of networks communities, that would explain the association between shortest path multiplicity and network communities as well.

In this work I investigate this hypothesis by means of numerical simulations and analytical calculations. In the Sec.~\ref{methods} I introduce different models of growing networks with local rules and the methods used to quantify their properties. In Sec.~\ref{random} I characterize random networks without local structures to set the baseline expectation. In Sec.~\ref{local} I characterize the networks generated by local rules. In Sec.~\ref{exact} I obtain the shortest path multiplicity exactly for two deterministic constructions, isolating the feature of the evolution rule that controls its growth.  In Sec.~\ref{real} I report data for real networks.  The concluding remarks are reported in Sec.~\ref{conclusions}.

\section{Methods}
\label{methods}

The computer code and the measured data related to these methods are available in a public repository~\cite{vazquez2026multipathdata}, at \href{https://github.com/av2atgh/ramsey_netcom/tree/master/local_multi_path_2026}{github.com/av2atgh/ramsey\_netcom}, in the directory \verb|local_multi_path_2026|. That directory contains the measured data together with the script \verb|generate_data_and_figs.py|, which reproduces Figs.~\ref{fig_randomized}--\ref{fig_real} from it.

\subsection{Local models}

Local search $LS(n, m)$. {\em Initial condition}. The network is started with two connected nodes. Evolution rule: A new node is added and a $m$-step random walk is performed from a randomly selected node in the current network. The new node is connected to all visited nodes, the starting node of the walk included. A node visited more than once by the walk receives a single link, so that the graph remains simple. This model has preferential attachment because the probability that a node is visited, beyond the entry node, is proportional to the current node degrees. Consequently it generates networks with a power law degree distribution. The $LS(n, m)$ networks have a
high clustering coefficient. At least 1 triangle, between the entry point and next neighbor visited, is formed at every node addition. For $m=1$ this model is equivalent to the triadic closure model \cite{growing_social_networks_jin01, communities_bianconi_2014}.

Duplication split $DS(n, q)$. {\em Initial condition}: The network is started with a cycle of four nodes, to enforce minimum degree 2 and minimum cycle length 4. {\em Evolution rule}: A new node $i$ is added to the network and a node in the current network is selected at random, node $j$. With probability $q$, $i$ becomes a duplicate of $j$ with links from $i$ to all neighbors of $j$, but not to $j$ itself. Otherwise, a link between $j$ and a randomly selected neighbor of $j$, node $k$, is split. The edge $(j, k)$ is removed and new edges $(i, j)$ and $(i, k)$ are created. This model has preferential attachment because, for the duplication rule, the probability that a node neighbor is duplicated is proportional to the node degree. The duplication rule does not make triangles and the split rule breaks triangles if they would exist. The main motif of this model is the formation of squares (cycles of length 4) between any two neighbors of the reference node $j$ and its duplicate $i$.

Bubble model $BB(n, L)$. {\em Initial condition}: The network is started with a cycle of $L+2$ nodes, to enforce minimum degree 2 and a cycle length $L+2$. {\em Evolution rule}: A chain of $L$ new nodes is added to the network. The two nodes at the chain ends are attached to the ends of an existing link, drawn uniformly at random among the current edges, creating a cycle of length $L+2$. The network grows in increments of $L$ nodes, and the sizes reported are those realized by the recursion. $BB(n, 1)$ is equivalent to an earlier Dorogovtsev-Mendes model where new nodes are connected to both ends of a randomly chosen link by two undirected links \cite{dorogotsev2001link}. The model has preferential attachment because the probability that a node is at the end of the selected link is proportional to its degree. The main network motifs are the cycles of length $L+2$ created by the evolution rule.

All these models have a finite Ramsey community number $r_\kappa$, the minimum graph size that guarantees the emergence of network communities with almost certainty \cite{vazquez2025local}.

\subsection{Barab\'asi-Albert model}

This is not a local model, but it is a standard reference. {\em Initial condition:} The network is started with a fully connected graph of $m+1$ nodes. {\em Evolution rule:} A new node $i$ is added to the network and connected to $m$ nodes in the current network. The nodes are selected with a probability proportional to their current degree.

\subsection{Network communities}

The network communities are inferred using the stochastic block model implemented in graph-tool (\verb|graph_tool.inference.minimize_blockmodel_dl|, with default parameters) \cite{peixoto_graph-tool_2014}. This stochastic block model finds the community structure with the minimum description length \cite{peixoto2024networkreconstructionminimumdescription}. In that sense, it gives as output the optimal number of communities $\kappa$ and the partition of the nodes into communities. The average of $\kappa$ is calculated from 100 realizations of network creation plus network community inference.

An important parameter is the correction for the network degree sequence, \verb| deg_corr: bool (optional, default: True)|. Some of the networks investigated have power law degree distributions, and therefore I use the degree-corrected version of the stochastic block model (default option). Similar results are obtained using the Infomap method implemented in the package with the same name \cite{mapequation2025software}, as previously shown \cite{vazquez2025local}.

\subsection{Network rewiring}

For network rewiring, I use the standard configuration model implemented with \verb|graph_tool.generation.random_rewire| with default parameters, except for \verb|n_iter|, set to 100. The configuration algorithm rewires the network links preserving the degree distribution \cite{configuration_method_park_2004}.

\subsection{Shortest path multiplicity}

The shortest path multiplicity $\mu_{ij}(G)$ between a pair of nodes $(i,j)$ in a graph $G$ is the number of distinct geodesics, i.e. paths of minimum length, connecting $i$ and $j$. For undirected graphs $\mu_{ij}=\mu_{ji}$, and $\mu_{ij}=0$ when $i$ and $j$ belong to different connected components. I use the graph-tool method \verb|graph_tool.count_shortest_paths(G, i, j)| with default parameters, or an equivalent breadth-first-search implementation for the larger networks.

The average shortest path multiplicity of a single graph is taken over the $n(n-1)$ ordered pairs of distinct nodes, the diagonal $i=j$ being excluded,
\begin{equation}
\mu(G) = \frac{1}{n(n-1)}\sum_{i\neq j}\mu_{ij}(G) = \binom{n}{2}^{-1}\sum_{i< j}\mu_{ij}(G),
\label{mu_def}
\end{equation}
the second equality following from the symmetry of $\mu_{ij}$. Disconnected pairs are kept in the sum, where they contribute zero. Equivalently, $\mu(G) = \psi(G)\, \mu_c(G)$, where $\psi(G)$ is the fraction of connected pairs and $\mu_c(G)$ the average restricted to connected pairs. The networks generated by the local rules are connected by construction, so that $\psi=1$ for them and the two averages coincide. In contrast, $\psi<1$ can occur after degree preserving rewiring and in the real networks.

In the terminology of Deng {\em et al} \cite{deng2026multipath}, $\mu_{ij}$ is the path multiplicity amount (PMA), the entries of the path multiplicity matrix, and the average \eqref{mu_def} is the path multiplicity index (PMI), for which those authors use the symbol $\Phi(G)$ and the same normalization by the $N(N-1)$ ordered pairs. The two definitions therefore coincide, $\mu(G)=\Phi(G)$, whenever the graph is connected. They differ only for disconnected graphs, which Deng {\em et al} handle by restricting the analysis to the giant connected component, whereas here the disconnected pairs are retained as zeros; the two conventions are related by the factor $\psi(G)$ above.

Given a graph generator $f_G$, I calculate the average shortest path multiplicity $\langle \mu\rangle$ as the average of $\mu(G)$ over 100 realizations of $G$.

\begin{figure}
\includegraphics[width=3.4in]{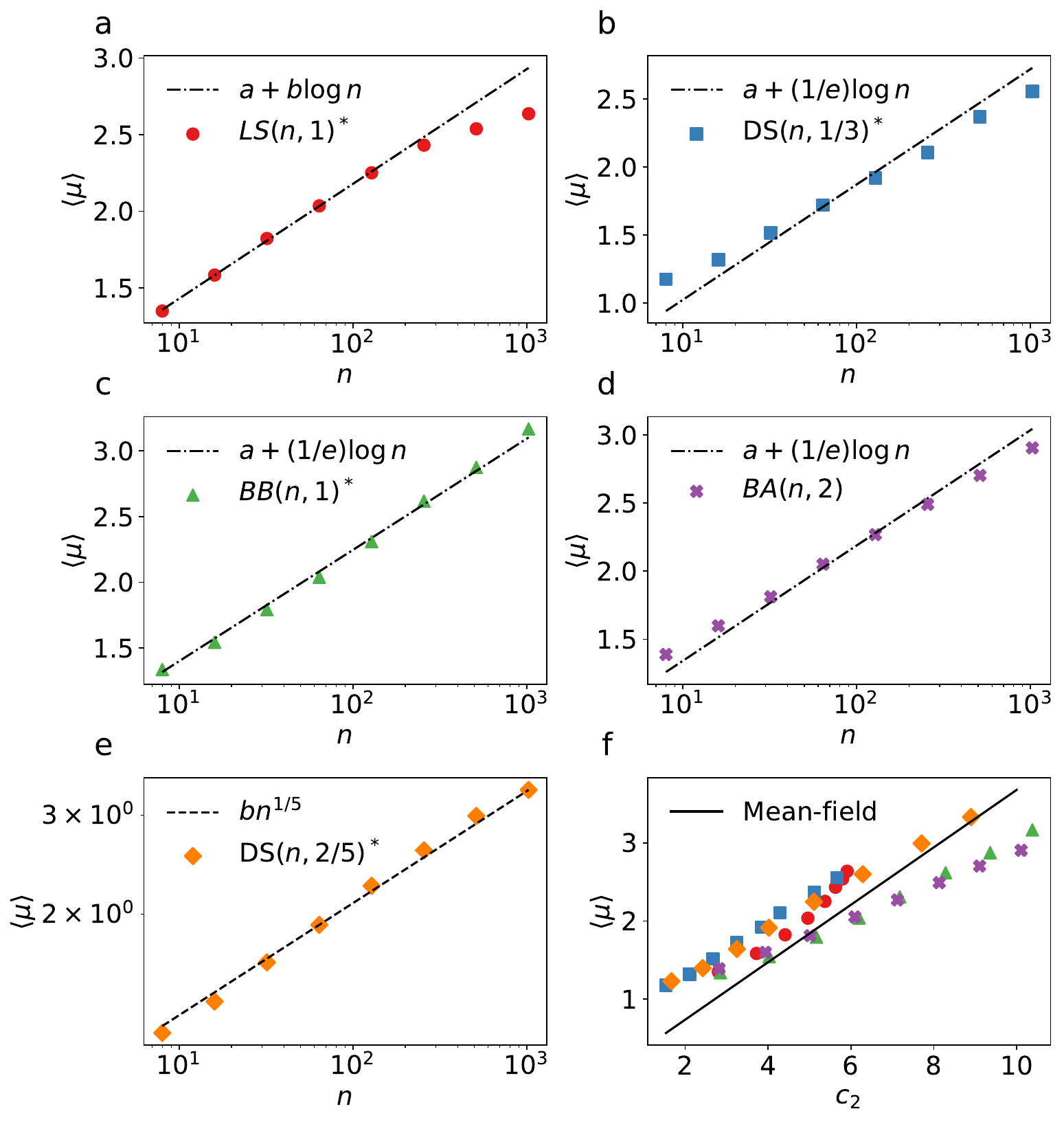}
\caption{Scaling between the average shortest path multiplicity $\langle \mu\rangle$ and the network size $n$ for degree preserving randomized networks ($*$) and for the Barab\'asi-Albert model without randomization. a) $LS(n,1)^*$, b) $DS(n,1/3)^*$, c) $BB(n,1)^*$, d) $BA(n,2)$, e) $DS(n,2/5)^*$. Network sizes $n=8$ to $1024$ in powers of two; each symbol is a mean over 100 realizations. The lines are the mean-field laws of Eqs.~\eqref{mu0}--\eqref{mu2} with only the additive constant $a$ adjusted to the data, the slope being fixed at the predicted value: $1/e$ in b-d ($\gamma=3$) and the exponent $\alpha=1/5$ in e ($\gamma=5/2$). A free power law fit in e gives $\alpha=0.21$ over the full range and $\alpha=0.20$ restricted to $n\geq64$, against the predicted $1/5$. In a the logarithmic law is fitted over $8\leq n\leq128$; the three largest sizes are excluded because $\langle\mu\rangle$ saturates there, as expected for $\gamma=5>3$. f) $\langle\mu\rangle$ vs the average excess degree $c_2$ for all five models, compared with the mean-field prediction $\langle\mu\rangle\sim c_2/e$ (line, no fitted parameter).}
\label{fig_randomized}
\end{figure}

\section{Random networks}
\label{random}

Using a mean field approach, I have estimated $\langle\mu\rangle$ for the Erd\"os-R\'enyi random graphs (Appendix A, Eq.~\eqref{muER})
\begin{equation}
\langle \mu\rangle_{ER} \sim \frac{c}{e} + \mathcal{O}(1), 
\end{equation}
where $c=\langle k\rangle$ is the average degree. For sparse graphs, $c$ is constant with increasing $n$ and so is $\langle \mu\rangle_{ER}$. In contrast, for minimally connected random graphs, $c\sim\ln n$, $\mu$ increases logarithmically with $n$. Finally, in the dense regime $c\sim n$, $\mu$ scales linearly with $n$, recapitulating a previous report by Dong {\em et al} \cite{dong2025multi}.

The same calculation can be adapted to networks with an arbitrary degree distribution $p_k$, resulting in 
\begin{equation}
\langle \mu\rangle \sim \frac{c_2}{e} + \mathcal{O}(1),
\label{muMF}
\end{equation}
where $c_2 = \langle k(k-1)\rangle/\langle k\rangle$ is the average excess degree  (Appendix A, Eq.~\eqref{muk}). In particular, for random graphs with a power law degree distribution $p_k\sim k^{-\gamma}$, there are three scenarios for $c_2$ (Appendix A, Eq. \eqref{c2N}). Using Eq.~\eqref{muMF} the $c_2$ scenarios are translated to
\begin{alignat}{2}
\langle \mu\rangle_{0}  &= a, &\text{for}\, \gamma>3,
\label{mu0}\\
\langle \mu\rangle_{I}  &= a + \frac{1}{e} \ln n, &\text{for}\,  \gamma=3,
\label{mu1}\\
\langle \mu\rangle_{II}  &= a + b n^{\alpha}, &\text{for}\, 2<\gamma<3,
\label{mu2}
\end{alignat}
where $a$ and $b$ are constants accounting for $n$-independent corrections and
\begin{equation}
\alpha = \frac{3}{\gamma} - 1.
\label{alpha}
\end{equation}

To investigate sparse random networks, we can use as starting point the local rule models and then apply the degree preserving link rewiring method. We have added the Barab\'asi-Albert model $BA(n,2)$ as well. Although the BA networks are not strictly random, their properties are close to their randomized versions preserving the degree distribution. We will add the notation $^*$ to indicate that the networks have been randomized.

For the randomized networks generated with the local search model $LS(n, 1)^*$, there is an apparent logarithmic growth up to $n\sim 100$, but saturating for larger network sizes (Fig. \ref{fig_randomized}a). The mean-field degree distribution has a power law tail with exponent $\gamma=5$ (Appendix A, Eq. \eqref{gammaLS} with $m=1$). Therefore, by Eq.~\eqref{mu0}, $\langle \mu\rangle$ is constant for large $n$. That is consistent with the saturation observed in the numerical results.

Next we focus on three models generating networks with the same expected degree distribution, with a tail exponent $\gamma=3$.  They are $BA(n,2)$, $DS(n, 1/3)^*$ and $BB(n, 1)^*$ (Appendix A). In this case, by Eq.~\eqref{mu1}, there is a logarithmic scaling with increasing the network size. The logarithmic scaling is confirmed by the numerical simulations (Fig. \ref{fig_randomized}b-d), albeit with small deviations from the mean-field slope $e^{-1}$.

Tapping into the case $\gamma<3$, we generated randomized networks using the duplication-split model $DS(n, 2/5)^*$. The expected degree distribution is a power law with tail exponent $\gamma= 5/2 =2.5$ (Appendix A, Eq. \eqref{gammaDS}, with $q=2/5$). In this case, by Eq.~\eqref{mu2}, we expect the power law scaling $\langle \mu\rangle \sim n^{1/5}$. The numerical simulations are consistent with the predicted power law scaling (Fig. \ref{fig_randomized}e).

Finally, all the data can be put together to challenge the mean field prediction that $\langle\mu\rangle\sim c_2/e$ (Appendix A, Eq. \eqref{muk}), where both $\langle\mu\rangle$ and $c_2$ are estimated numerically. The agreement with the mean-field result is not perfect, but overall the data for all models clusters near the mean-field line (Fig. \ref{fig_randomized}f).

\begin{figure}
\includegraphics[width=3.4in]{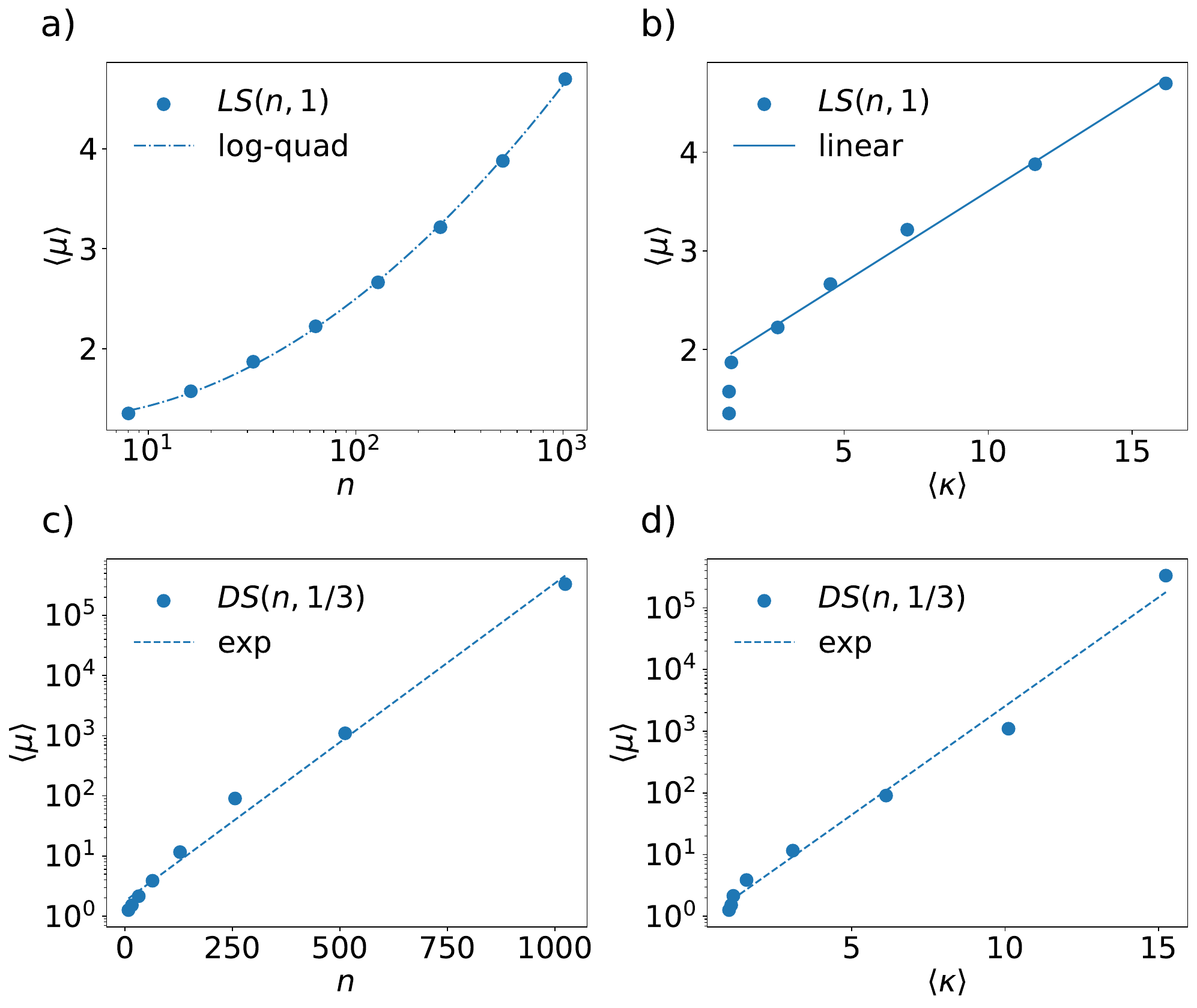}
\caption{Scaling between the average shortest path multiplicity $\langle \mu\rangle$ and the network size $n$ for networks generated by the local-search and duplication-split local rules, without randomization. a,b) $LS(n,1)$; c,d) $DS(n,1/3)$. Network sizes $n=8$ to $1024$ in powers of two; each symbol is a mean over 100 realizations of both the network generation and the community inference. The lines highlight the scaling specified in the legend: the log-quadratic law of Eq.~\eqref{mu_quad} in a, the exponential law of Eq.~\eqref{mu_exp} in c and d, and a linear law in b, each fitted over the full range of sizes shown, except in b and d where the sizes with $\langle\kappa\rangle=1$, for which no community is resolved, are excluded. Note that it takes a critical network size $r_\kappa$ to detect network communities. Therefore, the scaling between $\langle\mu\rangle$ vs $\langle\kappa\rangle$ manifests for the largest network sizes simulated.}
\label{fig_local}
\end{figure}

\section{Local network growth}
\label{local}

Now we investigate the networks generated by the local rules, without rewiring. Although these local rules do not contain any pre-defined community structure, the resulting networks have communities as determined by standard methods of network community detection \cite{vazquez2025local}. In fact, beyond a certain network size $r_{\kappa}$, the Ramsey community number, the observation of network communities is almost certain. Now we proceed to uncover the scaling of the average shortest path multiplicity with the network size and with the number of inferred communities.

In all models investigated $\langle\mu\rangle$ is an increasing function of $n$. Furthermore, for all the network models with local rules the  average number of inferred communities $\langle\kappa\rangle$ increases with increasing $n$ as well. As a consequence, we can investigate the implicit relation between $\langle\mu\rangle$ and $\langle\kappa\rangle$.

Figure \ref{fig_local}a and b report the data for the $LS(n, 1)$ model: node addition, link creation to a randomly chosen existing node, local search stopping at 1 step, and link addition to the reached node. For this model $\langle\mu\rangle$ increases faster than linear with increasing $\ln n$. A fit to the log-quadratic law
\begin{equation}
\langle\mu\rangle_{III} = a + b \ln n + c (\ln n)^2,
\label{mu_quad} 
\end{equation}
is consistent with the data points for the range of network sizes tested (Fig. \ref{fig_local}a, line). In turn, $\langle\mu\rangle$ exhibits a linear scaling with $\langle\kappa\rangle$ (Fig. \ref{fig_local}b). Note the log-quadratic growth is faster than the large $n$ saturation observed for the randomized  $LS(n, 1)^*$ networks (Fig. \ref{fig_randomized}a).

Figure \ref{fig_local}c and d report the data for the $DS(n, q=1/3)$ model: node addition and either node duplication with probability $q$, or link split otherwise. For this model $\langle\mu\rangle$ increases exponentially with $n$
\begin{equation}
\langle\mu\rangle_{IV} = a e^{ b n},
\label{mu_exp} 
\end{equation}
where $a>0$ and $b>0$ (Fig. \ref{fig_local}c). There is also an exponential scaling of $\langle\mu\rangle$ vs $\langle\kappa\rangle$ (Fig. \ref{fig_local}d). The exponential growth is faster than the logarithmic scaling observed for the randomized  $DS(n, 1/3)^*$ networks (Fig. \ref{fig_randomized}b).

\begin{figure}
\includegraphics[width=3.4in]{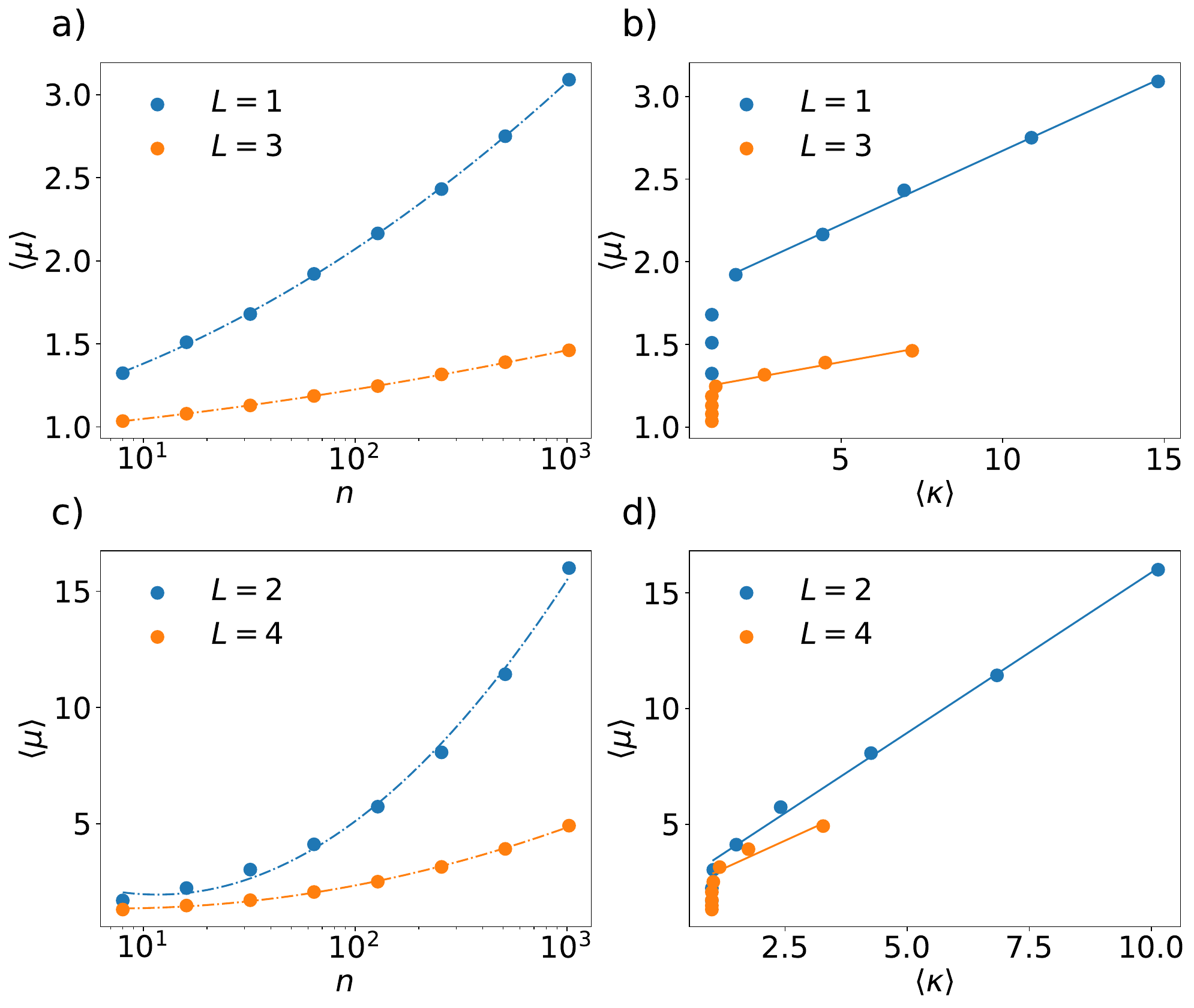}
\caption{Scaling between the average shortest path multiplicity $\langle \mu\rangle$ and the network size $n$ for networks generated by the $BB(n, L)$ model, without randomization. a,b) odd bubble lengths, $L=1$ and $L=3$; c,d) even bubble lengths, $L=2$ and $L=4$. Network sizes $n=8$ to $1024$; each symbol is a mean over 100 realizations. In a and c the lines are fits to the log-quadratic law of Eq.~\eqref{mu_quad} over the full range of sizes shown, and in b and d fits to a linear law, restricted to the sizes for which a community structure is resolved, $\langle\kappa\rangle>1$. Note that it takes a critical network size $r_\kappa$ to detect network communities. Therefore, the scaling between $\langle\mu\rangle$ vs $\langle\kappa\rangle$ manifests for the largest network sizes simulated.}
\label{fig_bubble}
\end{figure}

The differences between the local-search and duplication-split models may be related to the associated typical cycle length. Triadic closure rule in $LS(n, 1)$ creates cycles of length 3. There is only one shortest path between the nodes in a 3-nodes cycle. In contrast, the duplication rule creates cycles of length 4. Each node in a 4-cycle has two shortest paths to the opposite node two steps away. The duplication rule boosts the shortest path multiplicity. Note that this reasoning does not take into account the creation of multiple paths between existing nodes and the new node. Such additional contribution is responsible for the type I scaling for the $LS(n, 1)$ networks.

The bubble model $BB(n, L)$ is a great tool to investigate the cycle length dependency. At each network update, a chain of $L$ nodes is connected to the ends of a randomly chosen link, thus forming a cycle of length $L+2$. For $L=2k-1$ and $k=1, 2,\ldots$, the bubble model generates cycles of odd length. Otherwise, for $k=1, 2,\ldots$,  the cycles have even length.  The bubble networks have an expected degree distribution with power tail exponent $\gamma =3$ for $L=1$ and $\gamma>3$ for $L>1$ (Appendix A, Eq. \eqref{gammaBB}). For the randomized networks we would expect at most $\langle\mu\rangle \sim \ln n$, as corroborated for $BB(n, 1)^*$ in Fig. \ref{fig_randomized}c. In contrast, without randomization, the scaling is log-quadratic (Fig. \ref{fig_bubble}). While $\langle\mu\rangle$ reaches higher values for $L$ even than odd (Fig. \ref{fig_bubble}), the  existence of even length cycles is not sufficient for the exponential scaling.

\section{Exact results}
\label{exact}

The models considered so far can only be characterized numerically. I now turn to two deterministic constructions for which the shortest path multiplicity can be obtained exactly. Both are deterministic, all-edges versions of the bubble rule, and they differ in a single detail that turns out to control the scaling: whether the subdivided edge is retained or replaced.

\subsection{Deterministic bubble}
\label{deterministic}

Dorogovtsev, Goltsev and Mendes introduced a deterministic model for network growth with a recursive rule similar to the bubble model with $L=1$ \cite{dorogovtsev2002fractal}. The model is started at step $t=0$ with 3 connected nodes making a triangle. At each new step, new nodes are connected to both ends of every link in the current network, the original links being retained. This recursive rule can be extended to any value of $L$, obtaining the deterministic bubble model, here denoted by $BBD(n, L)$. Generation $t$ of $BBD(n,1)$ has
\begin{equation}
n_t = \frac{3^{t+1}+3}{2}
\label{nBBD}
\end{equation}
nodes and $3^{t+1}$ links, and a degree distribution with power law tail exponent $\gamma = 1 + \ln 3/\ln 2 \approx 2.585$ \cite{dorogovtsev2002fractal}.

For the deterministic networks themselves $\langle\mu\rangle$ can be calculated exactly. Let $A$, $B$ and $C$ be the vertices of the initial triangle, hereafter the corners, and let $G_t$ be the graph at step $t$. The construction is exactly renormalizable: $G_{t+1}$ consists of three copies of $G_t$ glued pairwise at the three corners, one copy carrying the pair $(A,B)$, one $(B,C)$ and one $(C,A)$. Three properties then make the multiplicity tractable.

(i) The geodesics between two nodes of the same copy are not affected by the rest of the graph. A path leaving a copy through one of its corners and re-entering through the other spends at least two steps outside, along $A$--$C$--$B$, whereas the two corners are adjacent inside the copy. The distances and multiplicities restricted to a copy are therefore those of $G_t$, and the pairs internal to a copy contribute exactly $3S_t$ to $S_{t+1}$, where $S_t = \sum_{i<j}\mu_{ij}$ is the total number of geodesics.

(ii) For a node $v$ of the copy carrying $(A,B)$, $d(v,C) = \min[d(v,A), d(v,B)] + 1$, because $C$ is adjacent to both $A$ and $B$ and lies outside that copy. A node is thus characterized by its two distances to the corners of its own copy, which differ by at most one, together with the corresponding numbers of geodesics.

(iii) Every path from a node $u$ to a node $v$ outside the copy of $u$ meets the copy boundary for the first time at exactly one corner. Writing $\tau_{uw}$ for the number of geodesics from $u$ to the corner $w$ whose first boundary contact is $w$,
\begin{equation}
\mu_{uv} = \sum_{w} \tau_{uw}\,\mu_{wv},
\label{firsthit}
\end{equation}
the sum running over the corners $w$ of the copy of $u$ that attain $\min_w [d(u,w) + d(w,v)]$. A first contact at a larger distance is never favorable, and $\tau_{uw}$ follows from the geodesic counts because the two corner distances differ by at most one. Because the first contact is unique, Eq.~\eqref{firsthit} requires no correction for paths visiting more than one corner.

Properties (i)--(iii) close a recursion for the joint distribution of corner distances and geodesic counts, and hence for $S_t$. The recursion involves $\mathcal{O}(t)$ distinct states and I have evaluated it exactly, reproducing the values obtained by direct enumeration of all node pairs for $t\leq7$ (Table \ref{tab:bbd}). The ratio $S_{t+1}/S_t$ converges to
\begin{equation}
\lambda = 10.049448332688990\ldots,
\label{lambdaBBD}
\end{equation}
the largest root of
\begin{equation}
\lambda^3 - 12\lambda^2 + 20\lambda - 4 = 0.
\label{cubicBBD}
\end{equation}
The number of nodes triples at every generation and the number of node pairs grows by a factor of nine, so that $\langle\mu\rangle$ grows by the factor $\lambda/9$ per generation. Converting the generation index into the network size with Eq.~\eqref{nBBD},
\begin{equation}
\langle\mu\rangle \simeq C n^{\alpha}, \qquad \alpha = \frac{\ln (\lambda/9)}{\ln 3},
\label{muBBD}
\end{equation}
with $\alpha = 0.100393163\ldots$ and $C = 1.166903298\ldots$. Equation~\eqref{muBBD} reproduces the exact values with a relative error of $10^{-6}$ at $t=12$ and $10^{-19}$ at $t=40$ (Table \ref{tab:bbd}).

\begin{table}[t]
\caption{Exact node count $n_t$, total number of geodesics $S_t$ and average shortest path multiplicity $\langle\mu\rangle_t = S_t/\binom{n_t}{2}$ for the deterministic bubble model $BBD(n,1)$, obtained from the renormalization of properties (i)--(iii) and checked against direct enumeration for $t\leq7$. The last column is the asymptotic law of Eq.~\eqref{muBBD}, with no adjustable parameter.}
\label{tab:bbd}
\begin{ruledtabular}
\begin{tabular}{lrrrr}
$t$ & $n_t$ & $S_t$ & $\langle\mu\rangle_t$ & $C n_t^\alpha$ \\
\colrule
1  & 6                    & 18                    & 1.200000   & 1.396870 \\
2  & 15                   & 150                   & 1.428571   & 1.531463 \\
3  & 42                   & 1422                  & 1.651568   & 1.698236 \\
4  & 123                  & 14046                 & 1.872051   & 1.891677 \\
5  & 366                  & 140442                & 2.102583   & 2.110527 \\
7  & 3282                 & $1.416\times10^{7}$   & 2.629224   & 2.630457 \\
12 & 797163               & $1.451\times10^{12}$  & 4.565716   & 4.565726 \\
40 & $1.82\times10^{19}$  & $1.666\times10^{40}$  & 100.161707 & 100.161707 \\
\end{tabular}
\end{ruledtabular}
\end{table}

The exact exponent can be contrasted with the expectation for a random network with the same degree sequence. Substituting $\gamma = 1 + \ln 3/\ln 2$ into Eq.~\eqref{mu2} gives $\alpha^* = -1 + 3/\gamma \approx 0.161$, larger than the exact $\alpha \approx 0.100$ obtained above. Destroying the correlations built by the deterministic rule, while preserving the degree sequence it generates, therefore {\em increases} the shortest path multiplicity.

The largest multiplicity in the graph grows faster than the average. I find that it is given by the Fibonacci numbers, $\max_{ij}\mu_{ij} = F_{t+2}$, verified up to $t=7$, which in terms of the network size is $n^{\ln\varphi/\ln 3} = n^{0.438}$, with $\varphi$ the golden ratio.

This model also makes explicit the origin of the correlation that motivates this work. The community structure of the pseudofractal web has been determined exactly \cite{vazquez2026pseudofractal}: the network is best described not by two communities but by a nested hierarchy, and the degree-corrected evidence is maximized at $\langle\kappa\rangle\sim\sqrt{n}$ communities of $\sim\sqrt{n}$ nodes each. Both the number of communities and the average shortest path multiplicity are therefore increasing functions of the network size, and the relation between them follows by eliminating $n$,
\begin{equation}
\langle\mu\rangle \sim \langle\kappa\rangle^{2\alpha} = \langle\kappa\rangle^{0.201}.
\label{mukappaBBD}
\end{equation}
A correlation between $\langle\mu\rangle$ and $\langle\kappa\rangle$ is thus obtained without either quantity acting on the other. Both are consequences of the same local rule, and the apparent dependency between them is induced by their common dependency on the network size.

\subsection{Diamond lattice}
\label{diamond}

In the deterministic bubble model the subdivided link is retained, so that the two ends of every bubble remain adjacent. I now consider the construction in which the link is instead replaced, the $(b,s)$ diamond hierarchical lattice~\cite{migdal1975phase, kadanoff1976notes}, and show that this single change alters the scaling qualitatively. Starting from a single bond between two poles $A$ and $B$ (generation $t=0$), every edge is replaced at each step by $b$ parallel paths of $s$ edges each. This is again a deterministic, all-edges version of the bubble rule $BB(n, L)$, which instead attaches one bubble of length $L=s$ to a single randomly chosen link per step. The standard diamond corresponds to $b=s=2$: every edge becomes two parallel two-edge paths. Generation $t$ has $n_t = \frac13(2\cdot4^{t}+4)$ nodes, so $n_t=2,4,12,44,\ldots$ for $t=0,1,2,3,\ldots$.

Because distinct copies of the previous generation meet only at corner vertices, the shortest path multiplicity is exactly renormalizable. Reading the construction backwards (deflation), $G_t$ is $b$ parallel branches, each $s$ copies of $G_{t-1}$ in series. Across a series of copies lengths add and multiplicities multiply, while across the $b$ parallel branches (all of equal length) the multiplicities add. Writing $D_t$ for the pole-to-pole distance and $P_t\equiv\mu_{AB}$ for the pole-to-pole multiplicity, this gives $D_t=s\,D_{t-1}=s^{t}$ and $P_t=b\,P_{t-1}^{\,s}$ with $P_0=1$. Taking $\log_b$ linearizes the recursion, $\log_b P_t = 1+s\log_b P_{t-1}$, and yields the closed form
\begin{equation}
P_t = b^{\,(s^{t}-1)/(s-1)} \;\xrightarrow{\,b=s=2\,}\; 2^{\,2^{t}-1}.
\label{eq:diamond_pole}
\end{equation}
For the diamond this reads $P_t=2^{\,D_t-1}$ with $D_t=2^{t}$: the number of geodesics between the antipodal poles is \emph{doubly exponential} in the generation. Equation~\eqref{eq:diamond_pole} was verified against direct construction of the lattice for $(b,s)\in\{(2,2),(3,2),(2,3),(4,2),(3,3)\}$.

The average over all node pairs is obtained from the same recursion. Every vertex lies on a pole-to-pole geodesic, $d(v,A)+d(v,B)=D_t$ (by induction on $t$: detours around the parallel branches are strictly longer), so a vertex is characterized by the single number $d(v,A)$. Splitting the pairs into those internal to one copy and those spanning two copies, the total geodesic count $S_t=\sum_{u<v}\mu_{uv}$ obeys $S_t=4S_{t-1}+\mathrm{Inter}_t$ for the diamond, where the inter-copy term is a corner-transit convolution of the two copies' pole-path counts, weighted by the factor $P_{t-1}$ of the shared corner traversal. Carrying out the recursion exactly (Table~\ref{tab:diamond}), the average multiplicity $\langle\mu\rangle_t=S_t/\binom{n_t}{2}$ follows the doubly-exponential law
\begin{equation}
\langle\mu\rangle_t \;\sim\; \kappa\,\frac{2^{\,2^{t}}}{4^{t}} \;\approx\; 2.1\,\frac{P_t}{n_t},
\qquad \kappa\to1.576 .
\label{eq:diamond_avg}
\end{equation}
That is, $\langle\mu\rangle_t=\Theta(P_t/n_t)$: the average tracks the pole multiplicity divided by the node count, up to an $O(1)$ prefactor $\kappa$ set by the fixed point of the corner-transit convolution. The sum is dominated by near-antipodal pairs, which inherit a geodesic count of order $P_{t-1}^{2}\sim P_t$, comparable to the pole value.

Equation~\eqref{eq:diamond_avg} is expressed in the generation index $t$. The scaling with the network size $n$ follows by inverting the node count. For large $t$, $n_t=\frac13(2\cdot4^{t}+4)\simeq\frac23\,4^{t}$, so that $4^{t}\simeq\frac32 n$ and hence $2^{t}\simeq\sqrt{3n/2}$. Substituting $4^{t}$ and $2^{2^{t}}=2^{\sqrt{3n/2}}$ into Eq.~\eqref{eq:diamond_avg}, the average multiplicity grows with the network size as a \emph{stretched exponential},
\begin{equation}
\langle\mu\rangle \;\approx\; a\,e^{\,b\sqrt{n}},
\label{eq:diamond_n}
\end{equation}
with
\begin{equation}
b=\sqrt{\tfrac32}\,\ln 2\approx0.849,
\qquad
a=\frac{2\kappa}{3}\approx1.05 .
\label{eq:diamond_ab}
\end{equation}
The rate $b$ is exact and $a$ is the amplitude of the leading term, $\ln\langle\mu\rangle\simeq b\sqrt{n}+\ln a$, the algebraic factor $1/n$ being subleading with respect to $\sqrt{n}$. Retaining it, the reduced form $\langle\mu\rangle\simeq a\,e^{\,b\sqrt{n}}/n$ reproduces the tabulated values to better than $1\%$ from $t=4$ on.

\begin{table}[t]
\caption{Exact node count $n_t$, pole distance $D_t=2^{t}$, pole multiplicity $P_t=2^{\,2^{t}-1}$, and average multiplicity $\langle\mu\rangle_t=S_t/\binom{n_t}{2}$ for the diamond ($b=s=2$), from the recursion and checked against direct construction. The last column shows the convergence of the prefactor $\kappa=\langle\mu\rangle_t\,4^{t}/2^{2^{t}}$ in Eq.~\eqref{eq:diamond_avg}.}
\label{tab:diamond}
\begin{ruledtabular}
\begin{tabular}{lrrrl}
$t$ & $n_t$ & $D_t$ & $P_t$ & $\langle\mu\rangle_t$ \quad ($\kappa$) \\
\colrule
1 & 4     & 2  & $2$              & $1.333$ \ \ (1.333) \\
2 & 12    & 4  & $8$              & $1.939$ \ \ (1.939) \\
3 & 44    & 8  & $128$            & $7.307$ \ \ (1.827) \\
4 & 172   & 16 & $32\,768$        & $418.1$ \ \ (1.633) \\
5 & 684   & 32 & $2.15\times10^{9}$  & $6.66\times10^{6}$ \ (1.589) \\
6 & 2732  & 64 & $9.22\times10^{18}$ & $7.11\times10^{15}$ (1.579) \\
\end{tabular}
\end{ruledtabular}
\end{table}

The diamond lattice thus realizes the fastest growth among the deterministic constructions solved here: doubly exponential in the generation, i.e.\ $\langle\mu\rangle_t\sim P_t/n_t$ with $P_t=2^{2^t-1}$, which in terms of the network size is the stretched exponential $\ln\langle\mu\rangle\sim\sqrt{n}$ of Eq.~\eqref{eq:diamond_n}. Its origin is the same $b$-fold parallel branching that sets the seam multiplicity of the local models. A geodesic threads $(s^{t}-1)/(s-1)$ successive scales and at each scale independently selects one of the $b$ parallel branches; the branch choices multiply, reproducing $P_t=b^{(s^{t}-1)/(s-1)}$. It is not, however, the fastest growth reported in this work. The duplication-split model reaches the exponential law of Eq.~\eqref{mu_exp}, and an exponential in the network size outgrows a stretched exponential. Deterministic, maximally symmetric parallel branching is thus a very efficient generator of multiple geodesics, but a stochastic rule that duplicates whole neighborhoods is more efficient still.

The comparison between the two exactly solvable constructions isolates what that ceiling requires. Both replace every link at every step, both are deterministic, and both are self-similar. They differ only in whether the subdivided link survives. In the deterministic bubble model it does, the two ends of every bubble remain adjacent, the pole-to-pole multiplicity stays equal to one, distances grow only logarithmically with the network size, and the multiplicity follows the power law of Eq.~\eqref{muBBD}. In the diamond lattice the link is removed, the poles separate as $D_t=s^t$, and the branch choices compound across scales. Retaining or removing one link per bubble is thus what separates a power law from a stretched exponential.

It is natural to ask whether this difference is captured by the degree correlations, as the real network data of Sec.~\ref{real} might suggest. It is not. Both constructions are disassortative, with Newman coefficients \cite{newman02a} $r=-0.084$ for the deterministic bubble at $n=9843$ and $r=-0.113$ for the diamond lattice at $n=10924$, and in both cases $r\rightarrow0$ as $n\rightarrow\infty$. Link replacement in fact makes the diamond lattice the more disassortative of the two, because every link then joins one of the newest nodes, of degree $b$, to an older one, whereas the deterministic bubble also retains links between two old nodes. Yet the deterministic bubble falls {\em below} its mean-field expectation, $\langle\mu\rangle/(c_2/e) = 0.27$ at $n=3282$, while the diamond lattice lies far {\em above} it, by 34 orders of magnitude at $n=10924$. The sign of the degree correlations therefore does not determine the sign of the deviation from the random expectation. The architecture built by the evolution rule does.

The relation between multiplicity and communities can again be made explicit. The community structure of the diamond lattice is also known exactly \cite{vazquez2026rgcrossing}: the Ramsey community number is a renormalization-group crossing, and the optimal partition is a hierarchy of $\langle\kappa\rangle\sim\sqrt{n}$ communities, the same square-root growth found for the pseudofractal web \cite{vazquez2026pseudofractal}. Eliminating $n$ between that result and Eq.~\eqref{eq:diamond_n},
\begin{equation}
\langle\mu\rangle \sim a\, e^{b^\prime \langle\kappa\rangle},
\label{mukappaDL}
\end{equation}
so that here the multiplicity grows exponentially with the number of communities, against the weak power law of Eq.~\eqref{mukappaBBD} for the deterministic bubble. The two exactly solvable cases therefore exhibit two very different $\langle\mu\rangle$ versus $\langle\kappa\rangle$ relations while sharing the same $\langle\kappa\rangle\sim\sqrt{n}$ community scaling. The correlation between multiplicity and communities is real in both, but its functional form is inherited from the evolution rule, not from any influence of the communities on the paths.

\section{Real networks}
\label{real}

\begin{figure}
\includegraphics[width=3.05in]{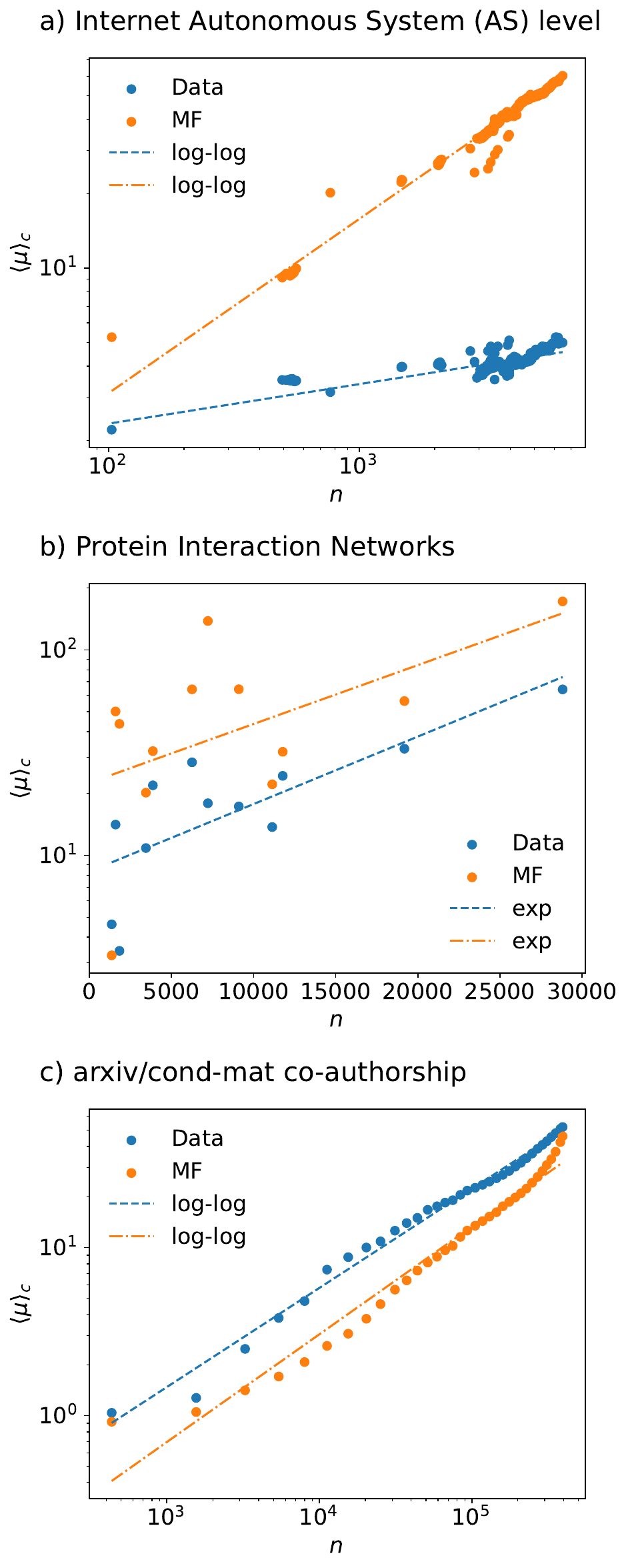}
\caption{Scaling of $\langle \mu \rangle_c$ vs $n$ in real networks (Data), together with the mean-field expectation $c_2/e$, Eq.~\eqref{muMF}, evaluated on the measured degree distribution with no fitted parameter (MF). a) Internet at the Autonomous System level, 733 daily snapshots; b) protein interaction networks of 12 organisms with more than 1000 proteins; c) the arxiv/cond-mat co-authorship network accumulated year by year. Each point is a single empirical network, and $\langle\mu\rangle_c=\langle\mu\rangle/\psi$ is the average restricted to connected pairs. The lines are fits over the full range, a power law in a and c and an exponential in b.}
\label{fig_real}
\end{figure}

To gain insights in the context of real networks, I have selected three systems with data at different stages of the network evolution. The Internet map at the Autonomous System (AS) level downloaded from \cite{snapnets}, where nodes are ASs and links are external connections between them. Protein interaction networks for different organisms \cite{stark2006biogrid}, excluding virus related networks and small networks with less than 1,000 proteins, where nodes are proteins and links are reported physical interactions between them. Finally, the co-authorship network of \verb|arxiv/cond-mat|, where nodes are authors and two authors are linked if they have co-authored a preprint \cite{cornell_arxiv_dataset}. The noise and non-stationarity of the data precludes any strong statements, but we can highlight some trends.

Figure \ref{fig_real} reports the average shortest path multiplicity vs the network size. For the real networks I report the average restricted to connected pairs, $\langle\mu\rangle_c = \langle\mu\rangle/\psi$, which is the convention adopted by Deng {\em et al} \cite{deng2026multipath}, and I write $\langle\mu\rangle_c$ throughout this section to keep the distinction explicit. This matters here because the empirical networks are not connected and their connected-pair fraction is itself a function of time: $\psi=1$ for the AS networks, but it ranges from 0.64 to 1.00 for the protein interaction networks and from 0.007 to 0.89 for the co-authorship networks, the smallest values corresponding to the earliest and sparsest samples. Reporting $\langle\mu\rangle_c$ ensures that the trends below reflect the multiplicity between reachable node pairs, and not the growth of the giant component. For the Internet AS networks, there is an apparent power law scaling between $\langle \mu \rangle_c$ and $n$ (Fig. \ref{fig_real}a, dashed line). The MF approximation, $c_2/e$, also exhibits an apparent power law scaling  (Fig. \ref{fig_real}a, dashed-dotted line), lying above the observed data. AS networks have a power law degree distribution $p_k\sim k^{-\gamma}$ with $\gamma\approx 2.1$ \cite{pastor-satorras01, internet_vazquez02}. The random networks with $\gamma<3$ also exhibit this power law scaling of $\langle \mu \rangle$ vs $n$ (Fig \ref{fig_randomized}e). However, the AS are not random, they have disassortative degree correlations \cite{pastor-satorras01, internet_vazquez02}, whereby the average degree of neighbor nodes decreases with increasing the node degree \cite{pastor-satorras01, internet_vazquez02} and the assortativity coefficient $r$ is negative \cite{newman02a}. Such star-like structure does not favor the formation of multiple shortest paths \cite{pastor-satorras01, internet_vazquez02}.

For the protein interaction networks there is an apparent exponential scaling between $\langle \mu \rangle_c$ and $n$ (Fig. \ref{fig_real}b, dashed line), although the data is too noisy. The MF approximation, also exhibits an apparent exponential scaling  (Fig. \ref{fig_real}b, dashed-dotted line), but lying above the observed data. Note that these networks exhibit the highest $\langle \mu \rangle_c$ values among the networks analyzed. Going from $\langle \mu \rangle_c\sim 3$ to $14$ for lower organisms on the lower $n$ range, to $\langle \mu \rangle_c\approx 64$ for the human network (largest network). The high values of $\langle \mu \rangle_c$ and the exponential trend match the behavior of the duplication-split model.

For the co-authorship networks there is an apparent power law scaling between $\langle \mu \rangle_c$ and $n$ (Fig. \ref{fig_real}c, dashed line). The MF approximation, also exhibits an apparent power law scaling  (Fig. \ref{fig_real}c, dashed-dotted line), but lying below the observed data. There are clear non-stationary patterns reflected in multiple inflection points. More prominent, at the last data point, corresponding with year 2026, the data and the MF expectation have close values. The current cond-mat co-authorship network has a shortest path multiplicity close to the random expectation. We note co-authorship networks exhibit assortative degree correlations, whereby the average degree of neighbor nodes increases with increasing the node degree \cite{vazquez03local} and the assortativity coefficient $r$ is positive \cite{newman02a}.

\section{Conclusions}
\label{conclusions}

In random networks with an arbitrary degree distribution, the average shortest path multiplicity scales linearly with the average excess degree, $\langle\mu\rangle \sim \langle k(k-1)\rangle/(\langle k\rangle e)$. When the degree distribution is a power law $p_k\sim k^{-\gamma}$, that implies the logarithmic scaling $\langle\mu\rangle_{I} = a + b\ln n$ for $\gamma =3$ and the power law scaling $\langle\mu\rangle_{II} \sim a n^\alpha$, $\alpha=-1+3/\gamma$, when $2<\gamma<3$. These predictions are partially supported by the numerical data. Some corrections to the mean-field calculations are needed.

In sparse networks generated by local evolution rules,  we observe two types of scaling of the average shortest path multiplicity $\langle\mu\rangle$ vs $n$. For most models there is a log-quadratic scaling $\langle\mu\rangle_{III}= a + b\ln n + c(\ln n)^2$. In contrast, for the duplication-split model, a faster exponential growth $\langle\mu\rangle_{IV} = a e^{bn}$ manifests.

Two deterministic constructions can be solved exactly, and together they identify what controls the growth. The deterministic bubble model gives a power law of the network size, the diamond hierarchical lattice a stretched exponential. The two differ in one detail only, whether the subdivided link is retained or removed. When it is retained the ends of every bubble stay adjacent and distances grow slowly; when it is removed the ends separate geometrically and the branch choices compound across scales. It is not the creation of cycles as such, but the separation of the nodes that the new cycle joins, that sets how fast the shortest path multiplicity can grow.

Since the local evolution rules induce (i) the formation of network communities and (ii) the average number of communities $\langle\kappa\rangle$ increases with increasing the network size, then we can investigate the scaling between $\langle\mu\rangle$ and $\langle\kappa\rangle$. It is linear for all models tested, except for the duplication-split model where an exponential dependency was observed.

Taken together, the results support the two-channel hypothesis raised in the introduction. Local rules are preferential, and the broad degree distribution that preferential attachment produces already makes $\langle\mu\rangle$ grow with the network size on its own. Local rules also close cycles, and that is what every stochastic local rule adds over its degree preserving randomization, which shares the same degree distribution. Neither channel on its own accounts for the observations. That is why the cause is named by the rule that produces both, rather than by either of its two consequences.

We could say local evolution rules and network communities are two sides of the same coin. With that view in mind, the statements that local evolution rules increase shortest path multiplicity and that network communities increase shortest path multiplicity  are equivalent. However, we should bear in mind that the local network evolution rules model the natural evolution of the systems they represent. In that regard, the local evolution rules are the driving mechanism and everything else follows. The evidence for that direction is a matched control: switching off the local rule by degree preserving randomization, at fixed network size, edge count and degree sequence, removes the excess multiplicity and returns the networks to the mean-field expectation $c_2/e$. The domain of this statement is that of growing networks, and that is not a restriction. Real networks are not drawn from a static ensemble. They are built node by node, and the mechanisms that build them are local: pages are copied from pages, references are found in the references we read, proteins gain connections when their partners are duplicated. A growing network with local rules is therefore not a special subclass, it is the generic representation of how real systems come to be, which is why the multiplicity of real networks is high in the first place. Local growth is plainly not necessary for a high shortest path multiplicity. A regular lattice is not grown by any such rule and already has many geodesics between distant nodes. The question is therefore not necessity but which mechanism is at work in the networks we actually observe, and those networks are grown node by node, by local rules. Mapping out the full boundary, with cycle-suppressed local rules and non-local modular or layered constructions, would sharpen the picture and is left for future work.

In real networks $\langle\mu\rangle$  increases with the network size, with different scalings depending on the system under consideration. Furthermore, although limited, the evidence points to a modulation by degree correlations. For networks with assortative degree correlations, the average shortest path multiplicity is above the MF approximation. In contrast, for networks with disassortative degree correlations, the average shortest path multiplicity is below the MF approximation. That association is not a rule, however, and the exactly solvable constructions show why. Both are disassortative, yet the deterministic bubble falls below the mean-field expectation while the diamond lattice lies far above it. The degree correlations are therefore something we checked and found insufficient: they do not by themselves predict the deviation from the random expectation, and the architecture built by the evolution rule has to be taken into account.

\begin{acknowledgments}
Nodes \& Links Ltd provided support in the form of salary for Alexei Vazquez, but did not have any additional role in the conceptualisation of the study, analysis, decision to publish, or preparation of the manuscript. Manuscript revisions and numerical checks were assisted by Claude Opus 4.8.
\end{acknowledgments}

\section*{Appendix A: Mean-field}
\label{MF}

\subsection*{Degree distribution}
\label{degree_dis}

We calculate the degree distribution using the rate equation approach \cite{degree_dist_krapivsky00}. It is a mean-field approximation because the expectation after an increase in network size is calculated from the expectations before the change, neglecting the stochastic nature of the node addition. For the models considered here, the expected number of nodes with degree $k$ after $n$ nodes has been added, $n_k(n)$, satisfies the recursive equation
\begin{eqnarray}
n_k(n+1) &=& n_k(n) + \frac{a + b(k-1)}{n} n_{k-1} - \frac{a+bk}{n} n_k \nonumber\\
&+& c \delta_{k2} + \frac{d}{n} n_k.
\end{eqnarray}
The last two terms account for the degree of the added node, which is either 2 or a duplicate of an existing node. Postulating the steady state solution $n_k(n) = np_k$ we obtain a recursion for $p_k$,
\begin{equation}
p_k = \frac{\eta + k -1}{\nu+k}p_{k-1} + \frac{c}{b} \frac{1}{\nu+k}\delta_{k2},
\label{pkpk}
\end{equation}
where
\begin{equation}
\eta = \frac{a}{b}, \quad \nu = \frac{1+a-d}{b}.
\end{equation}
From the recursion \eqref{pkpk} we derive the closed form
\begin{equation}
p_k = \frac{c}{b}\,
\frac{\Gamma(2+\nu)}{\Gamma(2+\eta)}\,
\frac{\Gamma(k+\eta)}{\Gamma(k+1+\nu)},
\label{pk}
\end{equation}
Proof by induction: First, it is valid for $k=2$. Second, evaluating $p_{k-1}$ in \eqref{pk} and substituting in \eqref{pkpk} we recover \eqref{pk} for $p_{k}$. Finally, the asymptotic behavior for $k\gg \max(\eta, \nu)$ is the power law $p_k\sim k^{-\gamma}$ with
\begin{equation}
\gamma = 1 + \frac{1-d}{b}.
\end{equation}
For $BA(2)$, $a=0$, $b = 1/2$, $c=1$ and $d=0$, resulting in
\begin{equation}
\gamma_{BA, m=2} = 3.
\label{gammaBA}
\end{equation}
For $LS(n, m)$, $a=1$, $b=m/\langle k\rangle = m/[2(m+1)]$, $c=1$, and $d=0$, resulting in
\begin{equation}
\gamma_{LS} = \frac{3m+2}{m}.
\label{gammaLS}
\end{equation}
For $DS(n, q)$, $a=0$, $b=q$, $c=1-q$, $d=q$, $\eta=0$ and $\nu=-1+1/q$, resulting in
\begin{equation}
\gamma_{DS}=\frac{1}{q}.
\label{gammaDS}
\end{equation}
For $BB(n, L)$, $a=0$, $b=2/\langle k\rangle = 1/(L+1)$, $c=1$, $d=0$, $\eta=0$ and $\nu=L+1$, resulting in
\begin{equation}
\gamma_{BB}=L+2.
\label{gammaBB}
\end{equation}
Note that $DS(n, q=(L+2)^{-1})$ and $BB(n, L)$ have the same degree distribution. Furthermore, $DS(n, 1/3)$, $BB(n, 1)$ and $BA(n, 2)$ all have the same degree distribution with exponent $\gamma=3$.

The excess degree scaling with the network size will depend on the $\gamma$ exponent. For $\gamma>3$, we expect $c_2\sim {\rm const.}$ for large $n$. For $\gamma \leq 3$, we first estimate the order of magnitude of the largest degree from the equation $p_{k_{\max}} =1/n$, resulting in $k_{\max}\sim n^{1/\gamma}$. Then, for $\gamma=3$, we expect $c_2 \sim \sum_{k=1}^{k_{\max}} \frac{1}{k} \sim \ln k_{\max} \sim \ln n$. In contrast, for $\gamma<3$, $c_2 \sim k_{\max}^{3-\gamma} \sim n^{-1 + 3/\gamma }$. Summarizing, for random networks with a power law degree distribution $p_k\sim k^{-\gamma}$, in the limit $n\rightarrow\infty$,
\begin{equation}
c_2 \sim \left\{
\begin{array}{ll}
{\rm const.} & {\rm for\,}\gamma>3,\\
\ln n & {\rm for\,}\gamma=3,\\
n^{-1 + 3/\gamma } & {\rm for\,} 2<\gamma<3.
\end{array}
\right.
\label{c2N}
\end{equation}

We can be more explicit for specific cases. For example, for $BA(n, 2)$, $DS(n, 1/3)$ and $BB(n, 1)$, the expected degree distribution is
\begin{equation}
p_k = A\frac{\Gamma(k)}{\Gamma(3+k)} = \frac{A}{(2+k)(1+k)k}.
\end{equation}
The first two moments are, for $n\rightarrow\infty$,
\begin{eqnarray}
\langle k\rangle &\sim& A \sum_{k=2}^{k_{\max}} \frac{1}{(2+k)(1+k)}, \nonumber\\
&\sim& A\left( \sum_{k=2}^{k_{\max}} \frac{1}{k+1} - \sum_{k=2}^{k_{\max}} \frac{1}{k+2} \right), \nonumber\\
&\sim& \frac{A}{3},
\end{eqnarray}
\begin{equation}
\langle k^2\rangle\sim A \sum_{k=1}^{k_{\max}} \frac{1}{k} \sim A \ln k_{\max} \sim \frac{A}{3} \ln n.
\end{equation}
Finally, putting both results together, 
\begin{equation}
c_{2, \gamma=3} \sim \ln n.
\label{c23}
\end{equation}

\subsection*{$\langle\mu\rangle$: Erd\H{o}s--R\'enyi graphs}

\noindent Here we calculate the average shortest path multiplicity for  an Erd\H{o}s--R\'enyi graph with $n$ nodes and link probability $p$, building on previous calculations in Refs.~\cite{Fronczak2004, distance_stats_baronchelli_2006, Blondel2007, SoleRibalta2019}. Throughout, $c= np$ denotes the mean degree, $D=\ln n /\ln c$ the typical distance between two nodes and we assume $c>1$. We calculate the average shortest path multiplicity using a mean-field approach. The paths between two nodes are assumed independent and their existence determined by $p$ and $n$. We denote by $m_d$ the number of potential paths between two nodes at a distance $d$, $M_d$ the number of realized paths between two nodes at a distance $d$ and $F_d$ the probability that that no path shorter than $d$ exists. The expected number of shortest paths between two vertices equals the expected number of length-$d$ paths $\langle M_d\rangle$, weighted by the probability that no shorter path exists,
\begin{equation}
\langle\mu\rangle = \sum_{d}\langle M_d\rangle\,F_{d-1}.
\label{mu_avg}
\end{equation}
A path of length $d$ between two fixed vertices uses $d-1$ ordered intermediate nodes drawn from the remaining $n-2$, so the number of candidate paths is
\begin{equation}
m_d = (n-2)(n-3)\cdots(n-d)\approx n^{\,d-1},
\label{md}
\end{equation}
for $d\ll n$. The distribution of the number of realized paths $M_d$ follows a binomial distribution with $m_d$ draws and probability of success $p^d$ (that all edges are present). Thus, the mean number of paths of length-$d$ is
\begin{equation}
\langle M_d\rangle = m_d\,p^{d} \approx \frac{(np)^{d}}{n} = c^{\,d-D}.
\label{Md}
\end{equation}
The probability that the paths are not present is $\Pr(M_d=0) =(1-p^d)^{m_d} \approx e^{ - \langle M_d\rangle }$, for $p\ll1$, and
\begin{equation}
F_d = \prod_{k=1}^d \Pr(M_k=0) = e^{-\sum_{k=1}^{d}\langle M_k\rangle}.
\label{eq:Fsum}
\end{equation}
The exponent is a geometric series of ratio $c>1$, dominated by its last term: $\sum_{k=1}^{d}\langle M_k\rangle \approx c^{d-D}$ for large $d$. Thus
\begin{equation}
F_d \approx  \exp\left( -c^{d-D} \right),
\label{Fd}
\end{equation}
recapitulating Eq. 14 in Ref.~\cite{Fronczak2004} and Eq. 3 in Ref.~\cite{Blondel2007}.

Substituting Eq.~\eqref{Md} and Eq.~\eqref{Fd} into \eqref{mu_avg} we obtain
\begin{equation}
\langle\mu\rangle \approx \sum_{d}c^{\,d-D}\,\exp\left( -c^{\,d-D-1} \right).
\label{mu_avg_final}
\end{equation}
The summand peaks sharply at $d=D+1$ resulting in
\begin{equation}
\langle\mu\rangle\approx\frac{c}{e}+O(1).
\label{muER}
\end{equation}
In the sparse regime ($c=np$ fixed, $n\to\infty$) Eq.~\eqref{muER} is constant in $n$. In contrast, in the minimally connected regime ($c=np=b\ln n$~\cite{ErdosRenyi1960, Bollobas2001}), Eq.~\eqref{muER} yields
\begin{equation}
\langle\mu\rangle\sim\frac{b}{e}\,\ln n.
\label{eq:connected}
\end{equation}

\section*{$\langle\mu\rangle$: Arbitrary degree distribution}

Here we calculate the average shortest path multiplicity for graphs with an arbitrary degree distribution $\pi_k$, mean degree $c_1 = \sum_k \pi_k k$ and excess degree $c_2 = \sum_k \pi_k k(k-1) / c_1$. I switch the notation from $p_k$ to $\pi_k$ because of the use of $p$ as probability of link existence in this context. The calculation follows the same steps as for an Erd\H{o}s--R\'enyi graph, with corrections in the probability that a link between two nodes exists.

Given a potential path of nodes, each node is selected at random from all nodes in the network (excluding those selected in previous positions) and labelled as first, second, ... and last. The probability that the first node is connected to the second is $p_1 = c_1/n$, with associated typical distance $D_1 = \ln n / \ln c_1$. From then on, we need to calculate the probability that a node at the end of a link is connected to a randomly selected node. The degree $k$ of the node at the link end is distributed as $k\pi_k/c_1$ and has excess degree $k-1$, excluding the node that we came from. The randomly selected node has degree $k^\prime$ with probability distribution $\pi_{k^\prime}$. The probability that the two nodes are connected is $kk^\prime/(nc_1)$ with mean value $p_2 = (c_2 c_1) / (n c_1) = c_2 / n$. The associated typical distance is $D_2 = \ln n / \ln c_2$. After these corrections, Eqs.~\eqref{Md} and \eqref{Fd} are replaced by
\begin{equation}
\langle M_d\rangle = \left\{
\begin{array}{ll}
m_1 p_1 = m_1 c_1^{1-D_1}, & {\rm if}\,d=1,\\
m_d p_1p_2^{d-1} \approx c_1 c_2^{d - D_2 - 1}, & {\rm otherwise};
\end{array}
\right.
\label{Mdk}
\end{equation}
\begin{eqnarray}
F_d &=& (1-p_1)^{m_1}\prod_{k=2}^{d}\left(1- p_1p_2^{k-1}\right)^{m_k}, \nonumber\\
&\approx& \exp\left( - \sum_{k=1}^d \langle M_k\rangle \right).
\label{Fdk}
\end{eqnarray}
The exponent in the latter equation is a geometric series of ratio $c_2>1$, dominated by its last term: $\sum_{k=1}^{d}\langle M_k\rangle \approx \langle M_d\rangle$ for large $d$. Thus
\begin{equation}
F_d \approx \exp\!\left[ - c_1 c_2^{d - D_2 - 1} \right] .
\label{Fdk1}
\end{equation}
Substituting Eq.~\eqref{Mdk} and \eqref{Fdk1} into Eq. \eqref{mu_avg} we obtain
\begin{equation}
\langle\mu\rangle \approx \sum_d c_1 c_2^{d - D_2 - 1} 
\exp\!\left[ - c_1 c_2^{d - D_2 - 2} \right].
\label{muk_avg_final}
\end{equation}
The summand peaks sharply at $c_1 c_2^{d - D_2 -2} = 1$, obtaining
\begin{equation}
\langle\mu\rangle \approx \frac{c_2}{e} + O(1).
\label{muk}
\end{equation}
For an Erd\H{o}s--R\'enyi graph, $c_2 = c_1 =c$, recovering Eq.~\eqref{muER}.

\bibliographystyle{apsrev4-1}


%

\end{document}